\documentclass[prb,twocolumn,showpacs,superscriptaddress,floatfix,amsmath]{revtex4}

\usepackage{graphicx}
\usepackage{dcolumn}

\begin{document}
\title{Density matrix renormalization group study of conjugated polymers
with transverse $\pi$-conjugation}
\author{Y. Yan and S. Mazumdar}

\affiliation{Department of Physics, University of Arizona, Tucson, AZ 85721}

\date{today}

\begin{abstract}
We report accurate numerical studies of excited state orderings in long hypothetical
$\pi$-conjugated oligomers in which the hydrogen atoms of trans-polyacetylene are
replaced with conjugated sidegroups, within modified Hubbard models. 
There exists a range of the bare Coulomb repulsion for which the excited state ordering
is conducive to photoluminescence in the substituted systems, even as this ordering
is opposite in the unsubstituted polyenes of the same lengths. Our work provides
motivation to study real $\pi$-conjugated  polymers with transverse conjugation
and small optical gaps.
\end{abstract}

\maketitle

Materials that luminesce in the infrared are of interest because of their potential
application in infrared lasers for telecommunications.
Designing $\pi$-conjugated polymers that can emit light in the infrared with
high quantum efficiency (QE) is an
intriguing scientific problem, as all known organic emissive systems with strong
photoluminescence (PL) emit either in the visible or
ultraviolet. Trans-polyacetylene (t-PA) 
with absorption threshold
at 1.6 eV, for example, is nonluminescent. This is a consequence of 
the Coulomb correlation between $\pi$-electrons, which drives the 
lowest two-photon state, the 2$^1$A$_g$, below the one-photon optical
exciton \cite{Hudson82}, the 1$^1$B$_u$. The optically pumped 1$^1$B$_u$ in this case
decays in ultrafast timescale to the 2$^1$A$_g$, radiative transition from
which is forbidden. Although single-walled carbon nanotubes (SWNTs) absorb in the  
infrared, their PL is extremely weak \cite{SWNT}. 
Theoretical work have found forbidden dark excitons below the
optical exciton in the SWNTs \cite{Hongbo04,Perebeinos04}, 
and it has been suggested that the reason for the
weak PL in these
systems is same as in t-PA \cite{Hongbo04}.
Strong PL in emissive $\pi$-conjugated polymers
such as poly-paraphenylene (PPP) and poly-paraphenylenevinylene (PPV)
is due to excited state ordering E(2$^1$A$_g$) $>$ E(1$^1$B$_u$)
in these systems [where E(...) is the energy of the state]. This reversed
excited state ordering (with respect to t-PA) is due to increased molecular
exciton character of the wavefunctions, which leads to confinement of the
particle and the hole in the excited 1$^1$B$_u$ state along the
backbone of the polymer. The confinement, as well as the reversed excited state
ordering are reproduced within {\it effective}
linear chain models for the emissive $\pi$-conjugated polymers, within which the
backbone phenyl groups are modeled by C-C bonds much stronger than
the standard ``double'' bond of t-PA \cite{Soos92,Soos93}. Since confinement
increases {\it both} E(1$^1$B$_u$) and E(2$^1$A$_g$), 
it is clear that the standard prescription
for light emission from $\pi$-conjugated polymers, viz., increasing effective bond
alternation \cite{Soos92,Soos93}, cannot lead to materials that emit in the infrared.

Mazumdar {\it et al.} have suggested that simultaneous small optical gap
and E(2$^1$A$_g$) $>$ E(1$^1$B$_u$) can be obtained by
substituting the hydrogen atoms of t-PA with conjugated side groups
\cite{Shukla99,Ghosh00}, a process that has been referred to as 
``site-substitution''. The original goal here was to obtain a theoretical
understanding of the high QE of the PL in polydiphenylacetylenes
(PDPAs) \cite{Gontia99,Hidayat00}.
The absorption and emission wavelengths
of the experimental PDPA systems are in the visible, but this is due to their
finite conjugation
lengths \cite{Tada97}. On the other hand, the experimental demonstration that PL
in PDPAs is 
from the backbone PA chain and not the trans-stilbene molecular
unit \cite{Hidayat00} confirmed the reversed excited state ordering in the PDPAs.
Theoretically, it was suggested that the phenyl substituents led to
electron delocalization away from the backbone C atoms 
into the transverse substituent groups,
causing a reduced
{\it effective} Hubbard repulsion between two $\pi$-electrons occupying the same 
backbone C atom \cite{Shukla99}. 
The smaller effective Hubbard repulsion, in turn, simultaneously 
lowered E(1$^1$B$_u$) and raised the relative E(2$^1$A$_g$). Alternatively, it was
also shown from explicit calculations that site-substitution 
caused not only transverse delocalization but also simultaneous longitudinal confinement
of the particle and the hole along the backbone in the excited 1$^1$B$_u$ state \cite{Ghosh00}.
The longitudinal particle-hole confinement then gives the reversed excited state
ordering, exactly as in polymers with large optical gaps \cite{Soos92,Soos93}.

Direct demonstration of E(2$^1$A$_g$) $>$ E(1$^1$B$_u$) in the
PDPAs
requires high order configuration interaction (CI) calculation that is difficult
because of the large number of C atoms in the unit cell \cite{Ghosh00}. 
Dallakyan {\it et al.}
\cite{Dallakyan03} have therefore performed ``proof-of-concept'' calculations for
a hypothetical prototype polymer, polydiethylenepolyacetylene (system {\bf [1]}
in Fig.~1), suggesting that PDPAs and other more complex systems can be modeled
by {\bf [1]}.
Correlated electron calculations for 
{\bf [1]} are simpler than for PDPAs due to the smaller number of C atoms in the
former. The number of C atoms even in {\bf [1]}, however,
increases rapidly with increasing number of unit cells. Exact or
full CI (FCI) calculations are possible only
for the two-unit oligomer of {\bf [1]}, and even quadruple CI (QCI) calculations 
become impossible
for more than three units. Calculations in reference \onlinecite{Dallakyan03} were therefore
performed within an approximate exciton model \cite{Chandross99}, within which only
the highest occupied and lowest unoccupied molecular orbitals (HOMO and LUMO)
of each hexatriene unit of {\bf [1]} were retained. This approximation allowed 
QCI calculations for up to five units of {\bf [1]}, and it was found within
dimerized Hubbard model calculations with an average hopping integral
of 2.4 eV that the critical 
``Hubbard $U$'' (hereafter $U_c$)
at which E(2$^1$A$_g$) becomes smaller than E(1$^1$B$_u$) 
(starting from $U$ = 0) is larger by $\sim$ 1 eV
in {\bf [1]} than in the unsubtituted polyene of the same length. The larger
$U_c$ in {\bf [1]} is sufficient as ``proof-of-concept'', as this 
indicates that there exists a range of values for the bare Coulomb repulsion for which
E(2$^1$A$_g$) $>$ E(1$^1$B$_u$) in the substituted oligomer even as the
opposite excited state ordering occurs in the unsubstituted system. In principle,
therefore, attaching side groups that are more extended 
than in {\bf [1]} can 
lead to systems with excited state ordering conducive to light emission.

The approximate nature of the exciton basis approach as well
as the severe chain length limitations are the obvious limitations of the previous
calculations \cite{Dallakyan03}. More importantly, the verification of the
original idea that enahanced particle-hole
delocalization into the sidegroups lead to confinement along the backbone
\cite{Ghosh00} requires that the computational technique be able to distinguish
between polymer {\bf [1]} and the slightly modified hypothetical polymers 
{\bf [2]} and {\bf [3]},
also included in Fig.~1. Within the proposed mechanism of reference
\onlinecite{Ghosh00}, for instance, the larger hopping integral corresponding to the
triple bond in the side group in {\bf [2]} should lead to a $U_c$  
even larger than in {\bf [1]}. Attaching an electron attracting CN group, as in
{\bf [3]}, should decrease the electron density on the backbone C atoms, and
enhance $U_c$ even more strongly \cite{Ghosh00}.
There is unfortunately no way to test these ideas precisely within the
exciton basis approach, as the  structural modifications in 
{\bf [2]} and {\bf [3]}, with respect to {\bf [1]},
change the HOMO and the LUMO of the unit cell weakly,
with the  changes distributed evenly over the other
MOs. 

\begin{figure}[tb]
\centerline{\resizebox{3.4in}{!}{\includegraphics{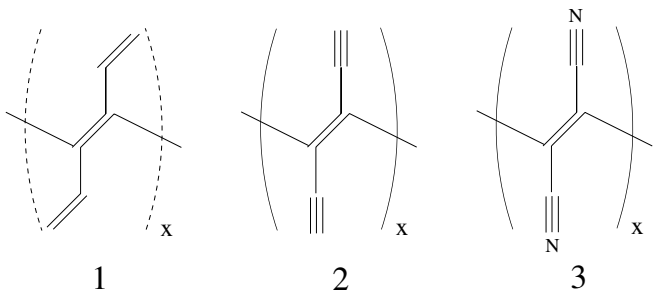}}}
\caption{Model hypothetical polymers with conjugated side groups
whose excited state orderings are of interest.
}
\label{polymers}
\end{figure}

In the present paper, we report the results of density matrix renormalization group
(DMRG) calculations of $U_c$ for all three systems {\bf [1], [2]} and {\bf [3]}.
The DMRG is a highly accurate numerical technique for correlated electron calculations
for long one dimensional systems  \cite{White} that has been extensively applied 
to determine the electronic structures of $\pi$-conjugated polymers
\cite{Ramasesha01,Barford05}. 
In order to investigate excited states,
we use the symmetrized DMRG 
(SDMRG) technique of Ramasesha {\it et al.} \cite{Ramasesha01}, which allows the
targeting of the lowest states of a given symmetry subspace.
In our calculations we exploit spatial,
electron-hole (for {\bf [1]} and {\bf [2]}) and
spin parity 
symmetries. Our wavefunctions are eigenstates of the total z-component of the spin,
S$_z$ and not the total spin S. 
This presents no problems in identifying the 1$^1$A$_g$ and the 1$^1$B$_u$, which 
are the lowest states with specific spatial and electron-hole
symmetries and total spin S = 0. In the
case of the 2$^1$A$_g$, however, care should be taken that the target S$_z$ = 0 state does
not belong to any of the S $>$ 0 states. We rule out the possibility of identifying
an incorrect excited state as the 2$^1$A$_g$ by explicitly calculating the transition
dipole moment between the target state and the 1$^1$B$_u$: dipole moments are nonzero 
only between states with the same total spin. 
The SDMRG technique allows the determination of
the energies of the 1$^1$A$_g$, the 1$^1$B$_u$ and the 2$^1$A$_g$ with high
accuracy upto 12 units (see below). We confirm that within the dimerized Hubbard model,
$U_c$ for {\bf [1]} is indeed larger than in the unsubstituted polyene of the
same chain length, as claimed before \cite{Dallakyan03}, although the difference
is smaller than in the previous approximate calculation. More importantly, though,
the $U_c$ for {\bf [2]} is substantially larger, while that for {\bf [3]} is larger
still, confirming our original hypothesis. We postpone discussions of the implications
of these results for real $\pi$-conjugated polymers until later.

As in reference \onlinecite{Dallakyan03}, we consider the modified Hubbard Hamiltonian,

\begin{subequations}
\begin{eqnarray}
H = H_{1e} + H_{ee} \\
H_{1e} = \sum_i \epsilon_in_i -\sum_{\langle ij \rangle,\sigma}t_{ij}c_{i,\sigma}^\dagger c_{j,\sigma} \\
H_{ee} = U\sum_{i}n_{i,\uparrow}n_{i,\downarrow}
\label{Hubbard}
\end{eqnarray}
\end{subequations}

\noindent In the above, 
$c_{i,\sigma}^\dagger$ creates a $\pi$-electron
of spin $\sigma$ on site $i$, $n_{i,\sigma}$ is the number of electrons with
spin $\sigma$ on site $i$, and
$\langle .. \rangle$ implies nearest neighbors.
$H_{1e}$ describes the one-electron site energies
and the nearest neighbor
hopping of electrons, and $H_{ee}$ consists of the electron-electron
(e-e) interaction within the Hubbard approximation. 
The hopping
integrals
$t_{ij}$ are taken to be $t_1$ = 2.4(1 -- $\delta$) eV and $t_2$ =
2.4(1 + $\delta$) eV, with $\delta$ = 0.07, corresponding to single and
double bonds, respectively. For the triple bonds in {\bf [2]} and
{\bf [3]} we choose $t_{ij}$ = 3 eV. These values of  $t_{ij}$ are
considered standard for $\pi$-conjugated systems within correlated electron
models \cite{Baeriswyl92}.
We
have ignored all Coulomb interactions other than the on-site repulsion $U$,
since
the 2A$_g$ - 1B$_u$ crossover is related to this interaction only, with the
spin-independent long range intersite Coulomb interactions merely modifying the
absolute magnitude of $U$ at which the crossover occurs. 
For the carbon-only
systems {\bf [1]} and {\bf [2]} in Fig.~1 we take all site energies $\epsilon_i$
= 0. For system {\bf [3]}, we have chosen the standard site energy 
of --3.0 eV for the nitrogen atoms \cite{Albert}. In principle, we should have 
also chosen a
different Hubbard $U$ for the nitrogen atoms, but since this difference is quite
small \cite{Albert}, and since our numerical procedure consists of varying the
Hubbard interaction, we have chosen the same $U$ for both carbon and nitrogen. 

The DMRG scheme that we use to calculate long chain behavior of the substituted
polymers of Fig.~1 is shown in Fig.~2. We add two atoms at each step of the
DMRG procedure. This approach proved to be better than adding single atoms at
each step.

\begin{figure}[tb]
\centerline{\resizebox{3.4in}{!}{\includegraphics{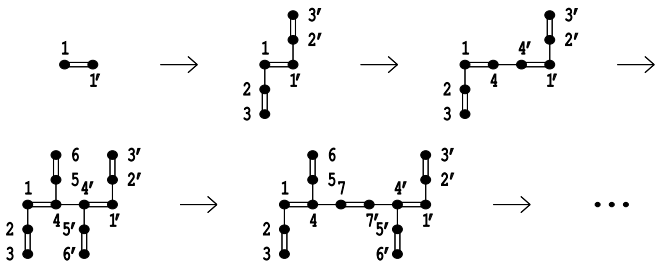}}}
  \caption{Schematic diagram of the building up of
the substituted polymer {\bf [1]}.
  }
\label{scheme}
\end{figure}

We have checked the accuracy of our calculations by comparing the 
DMRG energies at $U$ = 0 
of all three eigenstates we investigated
with their exact H\"uckel energies.
In Table 1, we have listed the DMRG and H\"uckel energies for the 1$^1$A$_g$, the
1$^1$B$_u$ and the 2$^1$A$_g$ for the 6- and 12-unit oligomers of {\bf [1]}, and
the 6-unit oligomers of {\bf [2]} and {\bf [3]}, respectively.
We find a high level of accuracy in all three absolute energies. The excitation energies
are slightly overestimated, with the overestimation of E(2$^1$A$_g$) 
being slightly larger than
that for E(1$^1$B$_u$) (for e.g., the overestimation for E(2$^1$A$_g$) of {\bf [3]} is
4\% while that for E(1$^1$B$_u$) is 1.2 \%.) We comment on this aspect  
following
the presentation of the numerical results, where we show that this does not change our
principal conclusion. Here we emphasize that
DMRG energies are least accurate in
the $U$ = 0 limit, where all wavefunctions are delocalized, and the
accuracy improves with increasing localization at larger $U$. 

%
\begin{table}[htb]
        \newcommand{\lb}[1]{\raisebox{-2mm}[0pt]{#1}}
        \newcommand{\rb}[1]{\raisebox{1.5mm}[0pt]{#1}}

  \caption{Comparison of DMRG energies at $U$=0 with H/"uckel energies 
(in units of $|t|$)
  }
  \centering
 \begin{tabular}{c|p{1.5cm}|c|c|c|c}
 \hline
 \lb{Polymer} & \centering Number of Units &  \lb{Method}  & \lb{1$Ag$}            & \lb{1$Bu$}  & \lb{2$Ag$}  \tabularnewline\hline

                        &                  & DMRG          & -45.31856     &       -44.94023       & -44.72728 \\
\rb{[1]}&\centering{\rb{6}}                     & Huckel        & -45.31860 &   -44.94138  & -44.73250 \\ \hline
                                                                        &                          & DMRG          & -91.08550 & -90.82209 & -90.69684 \\
\rb{[1]}&\centering{\rb{12}}            & Huckel  & -91.08582 &-90.82938        &-90.72156 \\ \hline

                        &                  & DMRG          & -49.25526 & -48.82161 & -48.56106 \\
\rb{[2]}&\centering{\rb{6}}                     & Huckel        & -49.25548 & -48.82795 & -48.58643 \\ \hline

                        &                          & DMRG          & -67.95981 & -67.52728 & -67.29174 \\
\rb{[3]}&\centering{\rb{6}}                     & Huckel        & -67.96001 & -67.53034 & -67.31854 \\ \hline

\end{tabular}
\end{table}

In Figs.~3(a) and (b) we have plotted the calculated DMRG
excitation energies E(1$^1$B$_u$) and
E(2$^1$A$_g$) with respect to the ground state energy 
for the 6- and 12-unit unsubstituted polyenes and oligomers of {\bf [1]}. These
results are to be compared with the approximate results for the 4- and 5-unit
oligomers \cite{Dallakyan03}. As in reference [\onlinecite{Dallakyan03}],
the excitation energies in the substituted polyenes are lower than the
unsubstituted systems. The initial increase in E(2$^1$A$_g$) with $U$ was seen
also in reference [\onlinecite{Dallakyan03}], as well as in earlier DMRG 
calculations of long
unsubstituted polyenes \cite{Ramasesha01}, where it was ascribed to the small
2$^1$A$_g$ - 1$^1$B$_u$ gap at $U$ = 0
in long chains. The increase persists in the 12-unit
oligomers for the range of $U$ we have studied, as the $U$ = 0
2$^1$A$_g$ - 1$^1$B$_u$ gap is quite small at this
chain length. The critical $U_c$ at which E(2$^1$A$_g$) $<$ E(1$^1$B$_u$) is indeed
larger in the substituted polyene {\bf [1]} than in the unsubstituted polyene,
as claimed before \cite{Dallakyan03}. However, the difference in the $U_c$ is
$\sim$ 0.5 eV as opposed to the $\sim$ 1.0 eV found previously.

\begin{figure}[tb]
\centerline{\resizebox{2.8in}{!}{\includegraphics{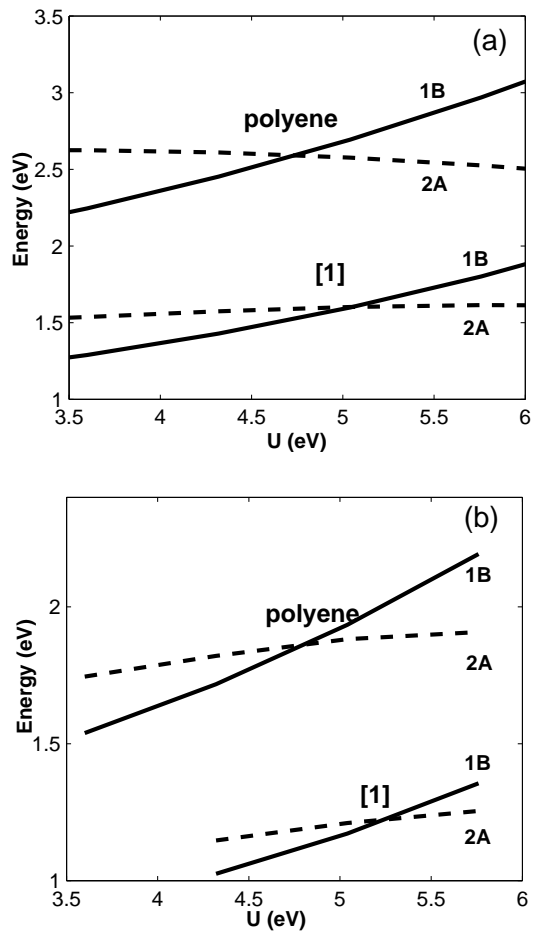}}}
\caption{(a) The excitation energies E(1$^1$B$_u$) and E(2$^1$A$_g$) for the 
unsubstituted polyene and the substituted system [1] of Fig.~1 for (a) 6-unit
, and (b) 12-unit oligomers.}
\label{substt}
\end{figure}

The critical $U_c$ are even larger in the oligomers of {\bf [2]} and {\bf [3]},
as seen in Fig.~4, where we have plotted E(1$^1$B$_u$) and E(2$^1$A$_g$) for all
three systems of Fig.~1. Detailed examination of the different behavior 
of the three systems provides indirect proof of our claim that site-substitution reduces
the effective Hubbard repulsion on the backbone C atoms. In particular, we draw
attention to the {\it qualitatively} different behavior of {\bf [2]} and {\bf [3]}. Both
E(1$^1$B$_u$) and E(2$^1$A$_g$) are larger in {\bf [2]}, relative to {\bf [1]}, but for
a given $U$, the increase in E(2$^1$A$_g$) is larger than that in E(1$^1$B$_u$). This
is the reason for larger $U_c$ in {\bf [2]}. This
feature is qualitatively similar to the behavior of these two energy gaps
in effective linear chains as the bond alternation is increased \cite{Soos92,Soos93}. 
The behavior of {\bf [3]}
is, however, completely different. E(2$^1$A$_g$) of {\bf [3]}, exactly as that of
{\bf [2]}, is larger than that of {\bf [1]} at all $U$. E(1$^1$B$_u$) of {\bf [3]}, on
the other hand, is larger than that of {\bf [1]} at $U$ = 0 but smaller than that of
{\bf [1]} for $U >$ 4.5 eV. It is precisely this feature of {\bf [3]}. i.e., 
relatively small E(1$^1$B$_u$) and large E(2$^1$A$_g$) at larger $U$ that gives the 
largest $U_c$ for this system. Such reduction of the optical gap with heteroatom 
substitution has also been observed in other theoretical work \cite{Albert,Lee}, but
the mechanism of the optical gap reduction here is quite different.
This behavior is precisely what is expected from the
effective Hubbard repulsion model \cite{Shukla99}. 
At small $U$ the optical
gap is determined by bond and site alternations, and E(1$^1$B$_u$) in
{\bf [3]} is larger than that of {\bf [1]} and is close to that of {\bf [2]} (see Table 1).
At moderate to large $U$ the ground state is predominantly covalent with all backbone 
C atoms singly occupied. The 1$^1$B$_u$ is ionic, with one vacant and one doubly occupied
C atom each. With electron attracting sidegroups,
the ``double occupancy'' in the 1$^1$B$_u$ is delocalized into the sidegroups, reducing the
effective Hubbard repulsion, and consequently the optical gap. The 2$^1$A$_g$ is covalent at 
large $U$, and hence its energy is unaffected (notice that the difference between the
E(2$^1$A$_g$) in {\bf [3]} and {\bf [1]} continues to be the same at small and large $U$
in Fig.~4).
Coming back now to the issue of overestimation of energies within the
DMRG procedure. $U_c$
for {\bf [3]} is so much larger than that of {\bf [1]} that even if the extent of
overestimation at $U$ = 0 is assumed to persist at $U \neq$ 0, and even if this
overestimation occurs only with E(2$^1$A$_g$) and not E(1$^1$B$_u$)
(in practice neither statement is true),
our statement regarding strong enhancement of $U_c$ by substitution with electron 
attracting sidegroups continues to be valid.

In conclusion, we have shown from accurate DMRG calculations that site substitution
with $\pi$-conjugated sidegroups can indeed simultaneously lower the optical gap 
and give excited state ordering
that is conducive the light emission. This is particularly true when the conjugated
sidegroup substituent is electron attracting. Although the present work involves 
hypothetical structures, we believe that they provide the motivation for examination 
of the excited
state ordering as well as the PL behavior of real 
``small bandgap'' polymers
with transverse $\pi$-conjugation, such as polyisothianaphthene and
poly(isonaphthothiophene), with optical gaps of 1.1 eV \cite{Kobayashi85} and 1.5 eV
\cite{Ikenoue90}, respectively. 
In contrast to the very intense studies of the wider optical gap polymers, 
these small gap polymers have received much less attention.
Good sample quality has also been elusive.
Theoretical work on these systems until now have focused on explaining the small optical gaps
and the competition between benzenoid and quinoid structures within one-electron theories
\cite{Kertesz90,Kertesz91}. Whether or not the observed small optical gaps can be explained
within many-electron theory, that is now established for the wider gap polymers,
is clearly of interest. 
We are currently
pursuing theoretical studies of these systems within the DMRG procedure to understand the
consequence of transverse $\pi$-conjugation on their optical gaps and excited state
ordering.

This work was partially supported by NSF-DMR-0406604. We acknowledge many useful
discussions on the DMRG procedure with S. Ramasesha from the I.I.Sc.,
Bangalore.

\begin{figure}[tb]
\centerline{\resizebox{2.8in}{!}{\includegraphics{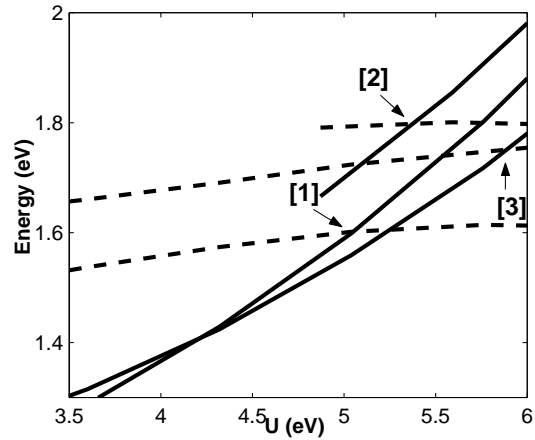}}}
\caption{The excitation energies E(1$^1$B$_u$) (solid  curves) and E(2$^1$A$_g$)
(dashed curves) for 6-unit oligomers of all 
three substituted systems of Fig.~1.}
\label{comparison}`
\end{figure}

\end{document}